\documentclass[12pt]{article}
\usepackage{epsf}
\def\beq{\begin{equation}}
\def\eeq{\end{equation}}
\def\bea{\begin{eqnarray}}
\def\eea{\end{eqnarray}}

\def\vel{\left|}
\def\ver{\right|}

\def\aga{\left\{}
\def\adr{\right\}}

\def\la{\langle}
\def\ra{\rangle}
\def\ba{\begin{array}}
\def\ea{\end{array}}
\def\ds{\displaystyle}
\title{ {\bf
$B\rightarrow K \tau^+ \tau^-$ decay in the general two Higgs doublet 
model }}
\author{\vspace{1cm}\\
        {\bf E. O. Iltan}
        \thanks{E-mail address:
        eiltan@heraklit.physics.metu.edu.tr}
 \\
        Physics Department, Middle East Technical University \\
        Ankara, Turkey\\}

\date{}

\begin{document}
\setlength{\baselineskip}{24pt}
\maketitle
\setlength{\baselineskip}{7mm}
\begin{abstract}
We study the branching ratio, $CP$-violating asymmetry, forward-backward 
asymmetry and the $CP$-violating asymmetry in the forward-backward asymmetry 
for the exclusive decay $B\rightarrow K \tau^+ \tau^-$ in the two Higgs 
doublet model with tree level flavor changing neutral currents (model III). 
We analyse the dependencies of these quantities on the neutral Higgs boson
contributions and the $CP$ parameter $sin\theta$ in the model III. We 
observe that to determine the neutral Higgs boson effects, the measurements 
of the forward-backward asymmetry and  the $CP$-violating asymmetry in the 
forward-backward asymmetry for the decay $B\rightarrow K \tau^+ \tau^-$ are 
promising.
\end{abstract} 
\thispagestyle{empty}
\newpage
\setcounter{page}{1}

\section{Introduction}
Rare B meson decays, induced by flavor changing neutral current (FCNC) 
$b\rightarrow s$ transition are the most promising research areas to test 
the Standard model (SM). Since these decays are induced at
loop level in the SM, a comprehensive information can be obtained for the 
more precise determination of the fundamental parameters, such as 
Cabbibo-Kobayashi-Maskawa (CKM) matrix elements, leptonic decay constants,
etc. Further, they shed light on the physics beyond the SM, such as 
two Higgs Doublet model (2HDM), Minimal Supersymmetric extension of the SM
(MSSM) \cite{Hewett}, etc. 

Among the rare B-decays, the ones which the SM predicts large branching ratio 
($Br$) become attractive since they are measurable in the near future, in the 
existing and forthcoming B-factories. The $B\rightarrow K l^+ l^-, 
(l=e,\,\mu,\,\tau)$ decay, induced by  $b\rightarrow s l^+l^-$ transition at 
the quark level, is one of the candidate. In the literature, there are
various experimental studies on this decay \cite{Babar1}-\cite{Babar2}. 
The $90\%$ C.L. upper limits of $Br(B\rightarrow K l^+ l^-)\,(l=e,\,\mu)$ 
have been obtained at the order of the magnitude as $10^{-6}$, close the SM
predictions.

In \cite{Hou} -\cite{Geng}, this transition has been investigated extensively 
in the SM, 2HDM . In these studies, the neutral Higgs boson (NHB) exchange 
diagrams are not taken into account since the lepton-lepton-Higgs vertices 
are proportional to the mass of the lepton underconsideration. However, for 
$l=\tau$ case, the mass $m_{\tau}$ can not be neglected since it is comparable 
with the $b$-quark mass and NHB exchange diagrams give considerable 
contributions to the physical quantities of such channels. In \cite{ASOK}, 
$B\rightarrow K \tau^+ \tau^-$ process is studied in the model II 2HDM and 
the NHB effects are taken into account. It is shown that the $Br$ ratio of 
the process is enhanced for large $tan\beta$ values and the NHB contributions 
become considerable. Recently, this decay has been  analysed and the forward
backward asymmetry has been studied in the constrained minimal supersymmetric 
SM \cite{Demir}. 

The forward-backward asymmetry $A_{FB}$ and the $CP$-violating 
asymmetry $A_{CP}$ are the physical quantities which provide information 
on the short distance contributions. $A_{FB}$ does not exist in the SM and
also in the 2HDM without NHB effects. However, with the addition of these 
effects, $A_{FB}$ is created and in the model II 2HDM, its numerical value 
increases with the increasing value of the vertex factor $tan\beta$ 
(see \cite{ASOK} for details). The sources of $A_{CP}$ are the complex CKM 
matrix elements or the Yukawa couplings appear beyond the SM. Since the $CP$ 
violation in the SM is negligible and no complex coupling exists in the 
model II (I) version of the 2HDM, one can go further and choose model III 
version of the 2HDM to get a measurable $A_{CP}$.

In our work, we study the exclusive $B\rightarrow K \tau^+ \tau^-$ decay 
in the model III. Since the $\tau$ lepton mass is comparable with $b$-quark
mass and the new Yukawa coupling $\xi_{N,\tau\tau}^D$, coming from the
vertices $\tau\, \tau \, h_0$ or $\tau \, \tau \, A_0$ can be large, we 
include the NHB diagrams and test the amount of their contributions. We 
calculate the $Br$ of the process and observe that it is sensitive to the 
NHB effects. Second, we get non-zero $A_{CP}$, at the order of the magnitude 
$10^{-2}$, since the Yukawa couplings in the model III can be taken as 
complex. $A_{FB}$ appears in the case that the NHB effects are non-zero and, 
therefore, we study its sensitivity to the Yukawa coupling 
$\xi_{N,\tau\tau}^D$ and the mass ratio, $\frac{m_{h_0}}{m_{A_0}}$, 
of the neutral Higgs bosons $h_0$, $A_0$. Finally we calculate the $CP$-
asymmetry in $A_{FB}$ $(A_{CP}(A_{FB}))$ and see that it is at the order 
of the magnitude $10^{-3}$. Similar to $A_{FB}$,  $A_{CP}(A_{FB})$ can 
exist if the NHB effects are non-zero and can be used for testing the 
contributions beyond the SM.  

Note that the theoretical analysis of exclusive decays is more complicated 
due to the hadronic form factors, which brings an uncertainity in 
the calculations. The calculation of the physical observables in the hadronic 
level needs non-perturbative methods to determine the matrix elements of the 
quark level effective Hamiltonian between the hadronic states. This problem 
has been studied in the framework of different approaches such as 
relativistic quark model by the light-front formalism \cite{Jaus}, chiral 
theory \cite{Casalbuoni}, three point QCD sum rules method \cite{Colangelo},  
effective heavy quark theory \cite{Roberts} and light cone QCD sum rules 
\cite{AOS}. 

The paper is organized as follows:
In Section 2,  we calculate the $Br$, $A_{CP}$, $A_{FB}$ and 
$A_{CP}(A_{FB})$ of the exclusive $B\rightarrow K \tau^+ \tau^-$ decay. 
Section 3 is devoted to the analysis of the dependencies of the physical 
quantities given above on the Yukawa coupling $\bar{\xi}_{N,\tau\tau}^{D}$, 
the ratio $\frac{m_{h_0}}{m_{A_0}}$ and the $CP$ parameter $sin\,\theta$. 
In appendix, we give a summary for the model III and the calculation of the 
matrix element for the inclusive $b\rightarrow s l^+ l^-$ decay in this 
model. Furthermore, we give the explicit forms of the form factors and the 
parametrizations used in the text. 
\section{The exclusive $B\rightarrow K l^+ l^-$ decay} 
The exclusive $B\rightarrow K l^+ l^-$ decay is induced by the inclusive 
$b\rightarrow s l^+ l^-$ process which has been studied in the literature 
extensively. Recently the $b\rightarrow s l^+ l^-$ decay has been handled with 
the addition the NHB effects in the framework of the general two Higgs doublet 
model \cite{gurer}. In the appendix  we give a summary of the model 
underconsideration and present the effective Hamiltonian which is used for the 
hadronic matrix elements.  

The calculation of the physical quantities like $Br$, $A_{FB}$, 
$A_{CP}$, etc., need the matrix elements $ \la K \vel \bar s \gamma_\mu 
(1+ \gamma_5) b \ver B \ra$, $ \la K \vel \bar s  (1+ \gamma_5) b \ver 
B \ra$ and $\la K \vel \bar s i \sigma_{\mu \nu} q^\nu (1+\gamma_5) b 
\ver B \ra$ and they read as \cite{ASOK}
\begin{eqnarray}
\la K \vel \bar s \gamma_\mu (1- \gamma_5) b \ver B \ra &=&
(p_B+p_K)_{\mu}\, f^+(q^2)+q_{\mu}\, f^- (q^2)\,\, , \nonumber \\  
\la K \vel \bar s i \sigma_{\mu \nu} q^\nu (1+\gamma_5) b \ver B \ra
&=& ( (p_B+p_K)_{\mu}\, q^2-q_{\mu} (m_B^2-m_K^2))\frac{f_T (q^2)}{m_B+m_K}
\,\, , \nonumber \\  
\la K \vel \bar s  (1+ \gamma_5) b \ver B \ra &=& \frac{1}{m_b}
((m_B^2-m_K^2) f^+(q^2)+q^2 f^- (q^2) )\,\, .   
\label{sand}
\end{eqnarray}
Here $p_{B}$ and $p_{K}$ are four momentum vectors of $B$ and $K$ mesons, 
$q=p_B-p_{K}$ is the momentum transfer. Using these form factors, the matrix 
element of the $B\rightarrow K l^+ l^-$ decay can be written as:
\begin{eqnarray}
{\cal M} &=& -\frac{G \alpha_{em}}{2 \sqrt 2 \pi} V_{tb} V_{ts}^*  
\Bigg\{  \left[ A p_{K \mu}+ B q_{\mu} \right] \bar \ell \gamma^\mu \ell 
+ \left[ C p_{K \mu}+ D q_{\mu} \right] \bar \ell \gamma^\mu\gamma_5 \ell
+ F_1 \bar \ell \ell + F_2 \bar \ell \gamma_5 \ell  \Bigg\}
\label{matr1}
\end{eqnarray}
where the functions $A$, $B$, $C$, $D$, $F_1$ and $F_2$ are given in  
Appendix B.  Using eq.(\ref{matr1}) and making the summation over final
lepton polarizations, the double differential decay rate is calculated as:  
\begin{eqnarray}
\frac{d \Gamma}{ds dz} &=& \frac{G^2 \alpha_{em}^2
|V_{tb} V_{ts}^*|^2 \, m_B }{2^{12} \pi^5 } \Bigg\{ v \sqrt{\lambda} 
( 
\frac{\lambda}{2} m_B^4 |A|^2+\frac{1}{2} |C|^2\, m_B^2 \,(\lambda\,
m_B^2+16\, m_l^2\, r)+2 |F_2|^2 m_B^2\, s 
\nonumber \\
&+& 8 Re(D^* F_2)\, m_B^2  m_l s + 8 |D|^2 m_B^2 m_l^2 s +4 Re(C^* F_2)
m_B^2 m_l (1-r-s)
\nonumber \\
&+& 8 Re(C^* D)\, m_B^2  m_l^2  (1-r-s) + 2 |F_1|^2 m_B^2 \,s\, v^2 +z 
(4 Re(A^* F_1) \sqrt{\lambda}\, m_B^2 \,m_l\, v)
\nonumber \\
&-&\frac{z^2}{2} \lambda \, m_B^4 \,v^2\, (|A|^2+|C|^2) 
) 
\Bigg\}
\label{dddW}
\end{eqnarray}
where $z=cos \theta$\,, $\theta$ is the angle between the momentum of $\ell$ 
lepton and that of $B$ meson in the center of mass frame of the lepton 
pair, $\lambda = 1+r^2+s^2 -2 r - 2 s - 2 r s$, $v=\sqrt{1-
\frac{4 m_l^2}{s\, m_B^2}}$, $r =\frac{\ds{m_{K}^2}}{\ds{m_B^2}}$ and 
$s=\frac{\ds{q^2}}{\ds{m_B^2}}$. 
In the light lepton case, namely $l=e,\mu$, the NHB effects are negligible
and new Wilson coefficients $C_{Q_1}$ and $C_{Q_2}$, appearing in the form
factors $F_1$ and $F_2$, almost vanish (see \cite{gurer} for details). 
However, for $\tau$ lepton, the NHB effects can give considerable 
contributions to the physical quantities $Br$, $A_{FB}$, $A_{CP}$, etc. Even 
if $A_{CP}$ is possible without these effects, they play the main role in the 
existence of $A_{FB}$. Further, $A_{CP} (A_{FB})$ exists when NHB effects are 
non-zero. Therefore, we concentrate on $A_{CP}(A_{FB})$ in addition to the 
quantities $A_{FB}$ and $A_{CP}$. Using the definitions 
\begin{eqnarray}
A_{FB} = \frac{\displaystyle{\int_0^1 dz \frac{d \Gamma}{dz} - \int_{-1}^0 dz
\frac{d \Gamma}{dz}}}{\displaystyle{\int_0^1 dz \frac{d \Gamma}{dz}+
\int_{-1}^0 dz\frac{d \Gamma}{dz}}} \, ,
\label{AFB}
\end{eqnarray}
\begin{eqnarray}
A_{CP} = \frac{\Gamma -\bar{\Gamma}}{\Gamma +\bar{\Gamma}}
\label{ACP}
\end{eqnarray}
we get 
\begin{eqnarray}
A_{FB}= \frac{\Phi}{\Omega}
\,\, , 
\label{AFB2}
\end{eqnarray}
and 
\begin{eqnarray}
A_{CP}= \frac{\int ds \Bigg{\{}2 A_2\, A_3\lambda^{\frac{3}{2}}\,v\, m_B^2 
\, (1-\frac{v^2}{3}) \, Im (\bar{\xi}^D_{N,bb})\Bigg{\}}}{\int ds \Delta}
\,\, , 
\label{ACP2}
\end{eqnarray}
where 
\begin{eqnarray}
\Phi &=& \int ds \Bigg{\{}\lambda \, v^2\, m_B^2 \,m_{\tau} (A_3\, 
F_{1}^{(2)} |\bar{\xi}^D_{N,bb}|^2 + A_2\, F_{1}^{(2)} Im (\bar{\xi}^D_{N,bb})
\nonumber \\ &+&  
(A_1\, F_{1}^{(2)}+A_3\, F_{1}^{(1)}) Re (\bar{\xi}^D_{N,bb})+ A_1
F_1^{(1)})\Bigg{\}}\,\, ,
\label{Phi}
\end{eqnarray}
and $\Omega$ is obtained from eq. (\ref{dddW}) by an integration over z
and s of the terms in the curly bracket. $\Delta$ in eq. (\ref{ACP2}) 
reads as
\begin{eqnarray}
\Delta &=& v\sqrt{\lambda} \Bigg{\{} m_B^2\, \lambda \, (A_1^2 + A_2^2 
+ C^2)\, (1-\frac{v^2}{3})+16\, m_{\tau}^2 (C D \sqrt{\lambda}+ 
(C^2 r+D^2 s))
\nonumber \\ &+& 
8\, m_{\tau}\,F_2^{(1)} ( \sqrt{\lambda\,}\, C + 2 \, s\, D)
+ 4\, s \,(|F_2|^2+v^2 |F_1|^2)\nonumber \\ &+& 
|\bar{\xi}^D_{N,bb}|^2  (m_B^2\, \lambda\, A_3^2\, (1-\frac{v^2}{3}) + 
4\,(F_1^{(2)})^2\, s\, v^2) + 2\, Re(\bar{\xi}^D_{N,bb})\, (
\lambda\, m_B^2\, A_1\, A_3 (1-\frac{v^2}{3}) + 4 F_1^{(1)}\, F_1^{(2)}\, 
s\, v^2 \nonumber \\ &+& 4\, m_{\tau} \sqrt{\lambda}\, C\, F_2^{(2)} + 
8\, m_{\tau}\, s\, D\, F_2^{(2)}) \Bigg{\}}
\label{delta}
\end{eqnarray}
The explicit forms of the functions $A_i,\,C,\,D,\,F_j^{(k)},\,\, 
i=1,2,3; \, j,k=1,2$ appearing in eqs. (\ref{Phi}) and
(\ref{delta}) are given in Appendix B. 

Finally $A_{CP}(A_{FB})$ can be defined as 
\begin{eqnarray}
A_{CP}(A_{FB}) = \frac{A_{FB} -\bar{A}_{FB}}{A_{FB} +\bar{A}_{FB}}
\label{ACPAFB}
\end{eqnarray}
where $A_{FB}$ is given in eq. (\ref{AFB2}) and $\bar{A}_{FB}$ can be
calculated by making the replacement $\bar{\xi}^D_{N,bb}\rightarrow
\bar{\xi}^{D *}_{N,bb}$ in $A_{FB}$. 

Notice that, during the calculations, we take into account only the second 
resonance for the LD effects coming from the reaction 
$b \rightarrow s \psi_i \rightarrow s \tau^{+}\tau^{-}$, where  $i=1,..,6$ 
and divide the integration region for $s$ into two parts : 
$\frac{4 m^2_{\tau}}{m^2_B}\leq s \leq \frac{(m_{\psi_2}-0.02)^2}{m^2_B}$ 
and $\frac{(m_{\psi_2}+0.02)^2}{m^2_B}\leq s \leq (1-\sqrt{r})^2$, where
$m_{\psi_2}=3.686\,GeV$ is the mass of the second resonance 
\section{Discussion}
The model III induces many free parameters, such as $\xi_{N,ij}^{U,D}$ 
where $i,j$ are flavor indices and they should be restricted using the 
experimental results. Now, we would like to present the restrictions we 
use for these parameters, in our numerical calculations. Since the neutral 
Higgs bosons, $h_0$ and $A_0$, can give a large contribution to the 
coefficient $C_7 $ which is in contradiction with the CLEO data 
\cite{cleo2}, 
\begin{eqnarray}
Br (B\rightarrow X_s\gamma)= (3.15\pm 0.35\pm 0.32)\, 10^{-4} \,\, ,
\label{br2}
\end{eqnarray}
the couplings $\bar{\xi}^{D}_{N,is}$($i=d,s,b)$ and $\bar{\xi}^{D}_{N,db}$ 
can be assumed as  negligible to be able to reach the conditions 
$\bar{\xi}^{D}_{N,bb} \,\bar{\xi}^{D}_{N,is} <<1$ and 
$\bar{\xi}^{D}_{N,db} \,\bar{\xi}^{D}_{N,ds} <<1$.
(see the appendix of \cite{alil2} for details).   
Using also the constraints coming from $\Delta F=2$ mixing, the $\rho$ 
parameter \cite{atwood} and the CLEO data we have:
\begin{eqnarray}
&  & \bar{\xi}_{N, tc} << \bar{\xi}^{U}_{N, tt} \,\, ,\nonumber \\ 
& & \bar{\xi}^{D}_{N, ib} \sim 0\,\, , 
\bar{\xi}^{D}_{N, ij} \sim 0, \,\, i,j=d,s \,\, quarks\, . \nonumber 
\end{eqnarray}
This assumption permits us to neglect the contributions coming from 
primed Wilson coefficients which are related with the chirality flipped
partners of the operator set (see \cite{gurer} for details) since the Yukawa 
vertices are combinations of  $\bar{\xi}^{D}_{N, sb}$ and 
$\bar{\xi}^{D}_{N, ss}$. Finally, we only take into account the Yukawa couplings 
$\bar{\xi}^{U}_{N, tt}$, $\bar{\xi}^{D}_{N, bb}$ and 
$\bar{\xi}^{D}_{N, \tau\tau}$. Notice that, for the coupling 
$\bar{\xi}^{D}_{N, \tau\tau}$, at first, we do not introduce any constraint
and we will try to predict an upper limit by usin the present experimental
measurements.  At this stage we introduce a new parameter $\theta$ with the 
expression 
\begin{eqnarray}   
\bar{\xi}^{U}_{N, tt}\,\bar{\xi}^{* D}_{N, bb}=
|\bar{\xi}^{U}_{N, tt}\,\bar{\xi}^{* D}_{N, bb}| e^{-i\theta} \, .
\label{theta}
\end{eqnarray}
Here, it is possible to take both $\bar{\xi}^{U}_{N,tt}$ and  
$\bar{\xi}^{D}_{N,bb}$ or any one of them complex. In our work, we choose 
$\bar{\xi}^{U}_{N,tt}$ as real and  $\bar{\xi}^{D}_{N,bb}$ as complex,
namely $\bar{\xi}^{D}_{N,bb}=|\bar{\xi}^{D}_{N,bb}|\, e^{i \theta}$. 
The phase angle $\theta$ leads to a substantial enhancement in neutron 
electric dipole moment and the experimental upper limit on neutron electric 
dipole moment $d_n<10^{-25}\hbox{e$\cdot$cm}$ thus places a upper bound on 
the couplings: $\frac{1}{m_t m_b} Im(\bar{\xi}^{U}_{N, tt}\,
\bar{\xi}^{* D}_{N, bb})< 1.0$ for $m_{H^\pm}\approx 200$ GeV \cite{david}.
 
In this section, we study the dependencies of 
the $Br$, $CP$ asymmetry $A_{CP}$, forward-backward asymmetry $A_{FB}$ and 
$CP$ asymmetry in forward-backward asymmetry $A_{CP}(A_{FB})$ of the decay 
$B\rightarrow K \tau^+ \tau^-$ on the selected parameters of the model III 
($\bar{\xi}^{D}_{N, \tau\tau}$,  $\frac{m_{h_0}}{m_{A_0}}$ and phase 
angle $\theta$). In our analysis we restrict $|C_7^{eff}|$ in the region 
$0.257 \leq |C_7^{eff}| \leq 0.439$, coming from CLEO measurement 
(see \cite{alil1} for details). With this restriction, an allowed region for 
the parameters  $\bar{\xi}^{U}_{N, tt}$, $\bar{\xi}^{D}_{N, bb}$ and 
$\theta$, is found. Throughout the numerical calculations, we respect this 
restriction, the constraint for the angle $\theta$ due to the experimental 
upper limit of neutron electric dipole moment and take 
$|r_{tb}|=|\frac{\bar{\xi}_{N,tt}^{U}}{\bar{\xi}_{N,bb}^{D}}| < 1$, the 
neutral Higgs mass $m_{H_0}=100\, GeV$, charged Higgs mass $m_{H^{\pm}}=
400\, GeV$, the scale $\mu=m_b$. Here, we also give the input values used
in the calculations, in Table (\ref{input}).  
\begin{table}[h]
        \begin{center}
        \begin{tabular}{|l|l|}
        \hline
        \multicolumn{1}{|c|}{Parameter} & 
                \multicolumn{1}{|c|}{Value}     \\
        \hline \hline
        $m_c$                   & $1.40$ (GeV) \\
        $m_b$                   & $4.80$ (GeV) \\
        $m_{\tau}$              & $1.78$ (GeV) \\
        $\alpha_{em}^{-1}$      & 129           \\
        $\lambda_t$            & 0.04 \\
        $\Gamma_{tot}(B_d)$    & $3.96 \cdot 10^{-13}$ (GeV)   \\
        $m_{B_d}$             & $5.28$ (GeV) \\
        $m_{t}$             & $175$ (GeV) \\
        $m_{W}$             & $80.26$ (GeV) \\
        $m_{Z}$             & $91.19$ (GeV) \\
        $\Lambda_{QCD}$             & $0.214$ (GeV) \\
        $\alpha_{s}(m_Z)$             & $0.117$  \\
        $sin\theta_W$             & $0.2325$  \\
        \hline
        \end{tabular}
        \end{center}
\caption{The values of the input parameters used in the numerical
          calculations.}
\label{input}
\end{table}

\begin{figure}[htb]
\vskip -3.0truein
\centering
\epsfxsize=6.8in
\leavevmode\epsffile{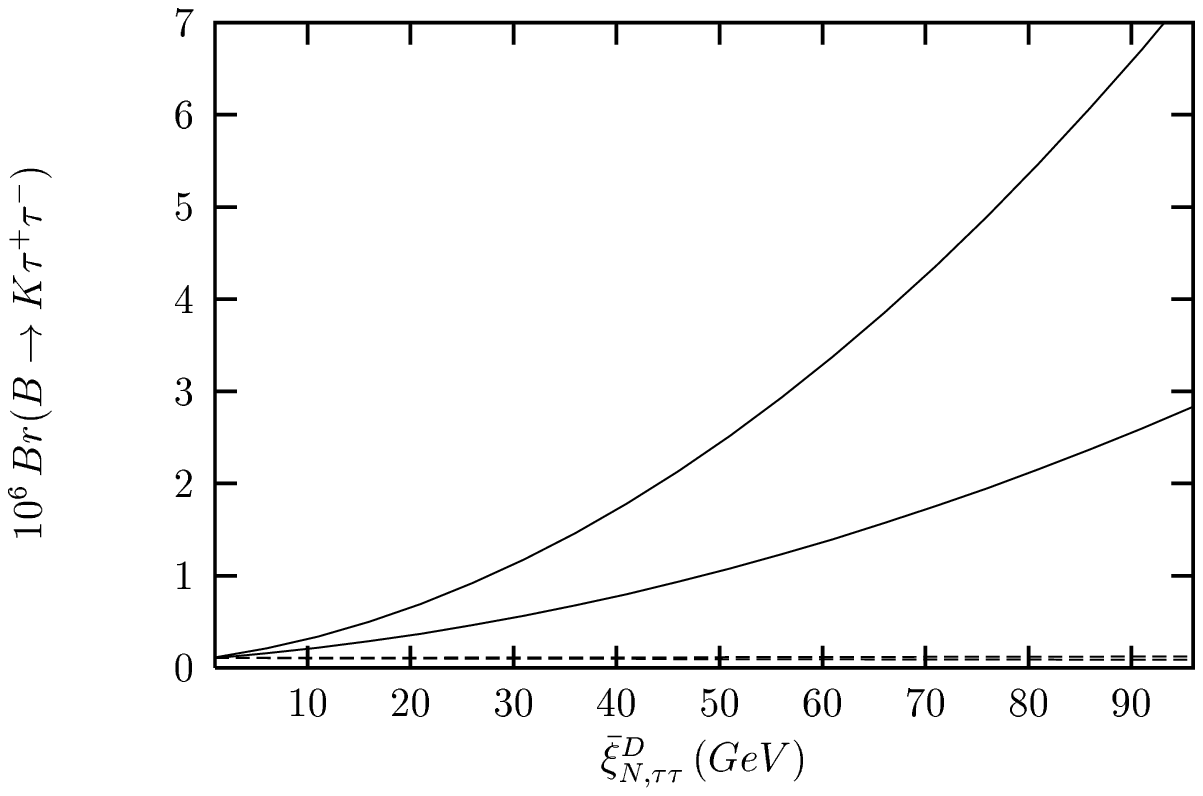}
\vskip -3.0truein
\caption[]{$Br$ as a function of $\bar{\xi}_{N,\tau\tau}^D$, for fixed 
$\bar{\xi}_{N,bb}^{D}=40\, m_b$, $m_{h_0}=70\, GeV$, 
$m_{A_0}=80\, GeV$ and $sin\theta =0$. Here $Br$ lies in the region 
bounded by solid (dashed) lines for $C_7^{eff} > 0$ ($C_7^{eff}<0$).} 
\label{Brktata}
\end{figure}

\begin{figure}[htb]
\vskip -3.0truein
\centering
\epsfxsize=6.8in
\leavevmode\epsffile{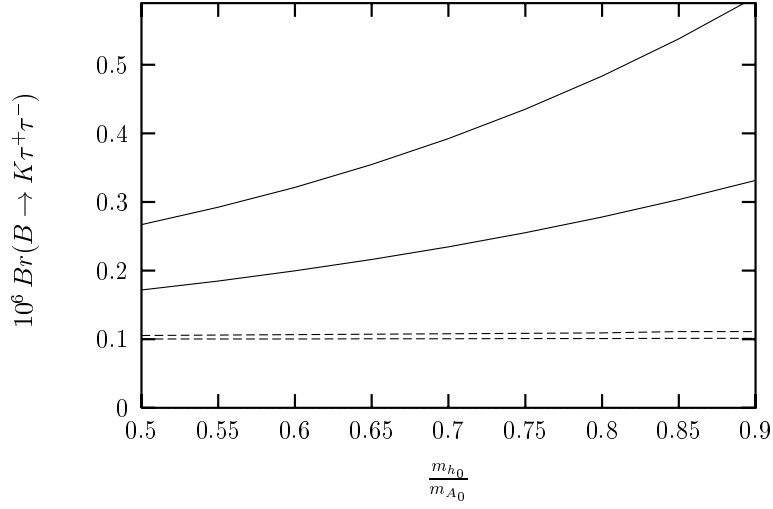}
\vskip -3.0truein
\caption[]{$Br$ as a function of $\frac{m_{h_0}}{m_{A_0}}$, for fixed 
$m_{A_0}=80\, GeV$, $\bar{\xi}_{N,bb}^{D}=40\, m_b$, 
$\bar{\xi}_{N,\tau\tau}^{D}=10\, m_{\tau}$ and $sin\theta =0$. Here $Br$ 
lies in the region bounded by solid (dashed) lines for $C_7^{eff} > 0$ 
($C_7^{eff}<0$).}
\label{Brh0A0}
\end{figure}

In  figs. \ref{Brktata} (\ref{Brh0A0}) we plot the $Br$ of the decay 
$B\rightarrow K \tau^+ \tau^-$ with respect to the Yukawa coupling 
$\bar{\xi}_{N,\tau\tau}^{D}$ (the ratio $\frac{m_{h_0}}{m_{A_0}}$)
for $\bar{\xi}_{N,bb}^{D}=40\, m_b$, $m_{h_0}=70\, GeV$, $m_{A_0}=80\, 
GeV$ ($\bar{\xi}_{N,\tau\tau}^{D}=10\,m_{\tau}$, $m_{A_0}=80\, GeV$). 
The $Br$ is restricted in the region between solid lines (dashed lines) 
for $C_7^{eff} > 0$ ($C_7^{eff} < 0$). Fig. \ref{Brktata} shows that $Br$ 
is sensitive to the NHB effects especially for $C_7^{eff} > 0$ case. For 
increasing values of $\bar{\xi}_{N,\tau\tau}^{D}$, $Br$ can take even two 
orders of magnitude larger values compared to the case where no NHB is taken 
into account. From this figure, it is possible to predict the upper limit 
of the coupling $\bar{\xi}_{N,\tau\tau}^{D}$, 
$\bar{\xi}_{N,\tau\tau}^{D} < \,20\, GeV$, respecting the experimental
upper limit, $Br ( B\rightarrow K l^+ l^-) < 0.5 \times 10^{-6}$ , 
$(l=e,\mu)$ \cite{Babar2}, with the assumption that the 
$Br (B\rightarrow K \tau^+ \tau^- )$ is not so much different than the 
previous one. For $C^{eff}_7 < 0$ the possible numerical values lie near 
$10^{-7}$ and the $Br$ is not sensitive to the NHB effects. 
Note that, the $Br$ in the SM is $1.06\, 10^{-7}$ and in the model III 
without NHB effects are between upper and lower  limits 
$(1.05-1.08)\,10^{-7}$  ($(0.95-0.97)\,10^{-7}$) for $C_7^{eff} > 0$ 
($C_7^{eff} < 0$). As shown in Fig. \ref{Brh0A0}, the $Br$ is also sensitive 
to the ratio $\frac{m_{h_0}}{m_{A_0}}$ for $C_7^{eff} > 0$ case and it 
increases for the larger values of the ratio. Furtermore, the experimental 
value of the $Br ( B\rightarrow K l^+ l^-)$ shows that the mass values  of 
the neutral Higgs bosons $h_0$ and $A_0$ are not far. 

\begin{figure}[htb]
\vskip -3.0truein
\centering
\epsfxsize=6.8in
\leavevmode\epsffile{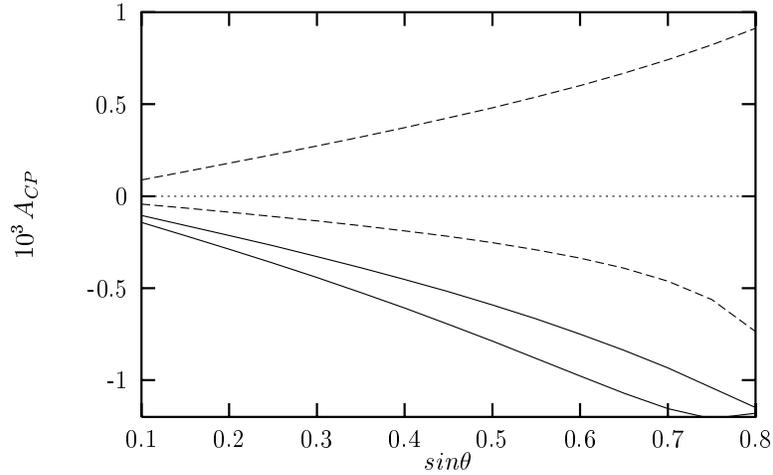}
\vskip -3.0truein
\caption[]{$A_{CP}$ as a function of  $sin\theta$ , for fixed 
$\bar{\xi}_{N,\tau\tau}^D=10\, m_{\tau}$, $\bar{\xi}_{N,bb}^{D}=40\, m_b$, 
$m_{h_0}=70\, GeV$, and $m_{A_0}=80\, GeV$. Here $A_{CP}$ lies in the region 
bounded by solid (dashed) lines for $C_7^{eff} > 0$ ($C_7^{eff}<0$).}
\label{ACPsin}
\end{figure}

\begin{figure}[htb]
\vskip -3.0truein
\centering
\epsfxsize=6.8in
\leavevmode\epsffile{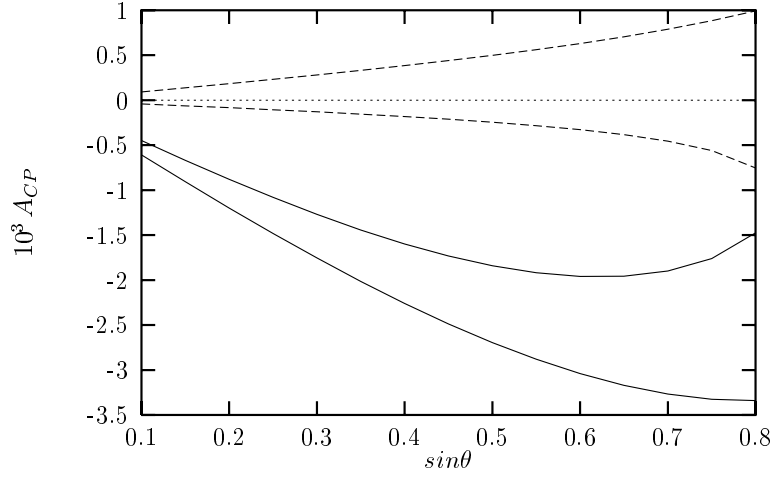}
\vskip -3.0truein
\caption[]{The same as Fig. \ref{ACPsin} but without NHB effects.}
\label{ACP0sin}
\end{figure}

Fig. \ref{ACPsin} is devoted to the $sin\theta$ dependence of $A_{CP}$
including NHB effects. Here, $A_{CP}$ is restricted in the region between 
solid lines (dashed lines) for $C_7^{eff} > 0$ ($C_7^{eff} < 0$). It is at 
the order of the magnitude $10^{-3}$ and increases with the increasing 
values of the parameter $sin\theta$ as it should be. For $C_7^{eff} > 0$, 
the possible values of $A_{CP}$ have the same sign (here minus), however 
for $C_7^{eff} < 0$ it can vanish or it can have both signs. We also 
present the $sin\theta$ dependence of $A_{CP}$ without NHB effects,  
in Fig. \ref{ACP0sin}. For this case $A_{CP}$ is greater as a magnitude 
and the restriction region is larger compared to the previous one,
expecially for $C_7^{eff} > 0$. As it can be seen from eq. (\ref{ACP2}),
the addition of NHB effects reduces the magnitude of the $A_{CP}$ since 
NHB contributions enter into expression in the denominator but not 
in the numerator.   

\begin{figure}[htb]
\vskip -3.0truein
\centering
\epsfxsize=6.8in
\leavevmode\epsffile{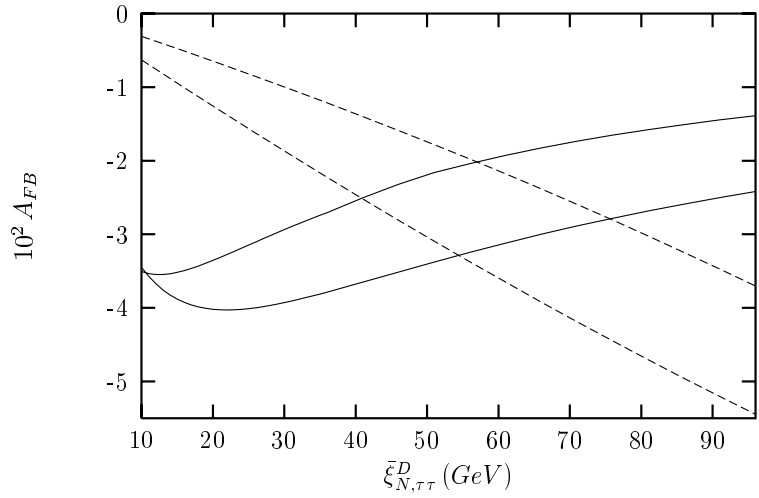}
\vskip -3.0truein
\caption[]{The same as Fig. \ref{Brktata}, but for $A_{FB}$ as a function of 
$\bar{\xi}_{N,\tau\tau}^{D}$.}
\label{AFBktata}
\end{figure}

\begin{figure}[htb]
\vskip -3.0truein
\centering
\epsfxsize=6.8in
\leavevmode\epsffile{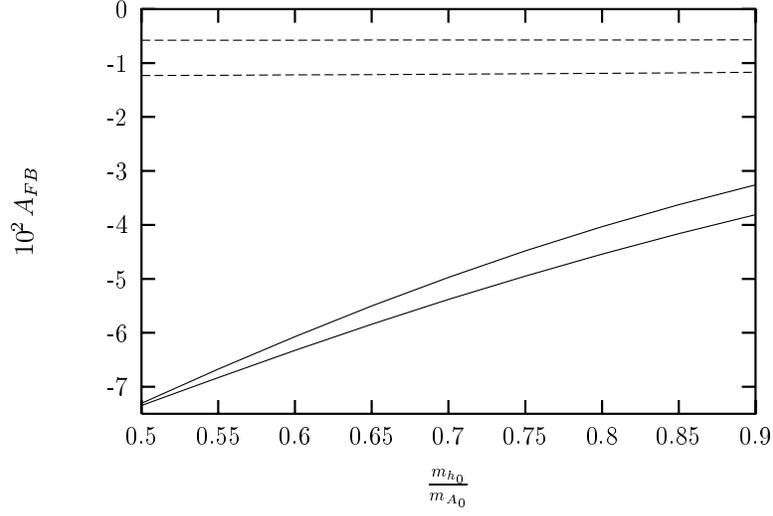}
\vskip -3.0truein
\caption[]{The same as Fig. \ref{Brh0A0}, but for $A_{FB}$ as a function of 
$\frac{m_{h_0}}{m_{A_0}}$.}
\label{AFBh0A0}
\end{figure}

In Figs. \ref{AFBktata} (\ref{AFBh0A0}), we plot the $A_{FB}$ of the decay 
$B\rightarrow K \tau^+ \tau^-$ with respect to the Yukawa coupling 
$\bar{\xi}_{N,\tau\tau}^{D}$  (the ratio $\frac{m_{h_0}}{m_{A_0}}$) for 
$\bar{\xi}_{N,bb}^{D}=40\, m_b$, $m_{h_0}=70\, GeV, m_{A_0}=80\, GeV$ 
($\bar{\xi}_{N,\tau\tau}^{D}=10\,m_{\tau}$, $m_{A_0}=80\, GeV$ ). $A_{FB}$ 
is restricted in the region between solid lines (dashed lines) for 
$C_7^{eff} > 0$ ($C_7^{eff} < 0$). Since $A_{FB}$ appears only with the 
NHB effects, it is a good candidate for testing the existence 
of them. $A_{FB}$ is at the order of the magnitude $10^{-2}$ and 
increases with the increasing values of $\bar{\xi}_{N,\tau\tau}^{D}$ for 
$C_7^{eff} < 0$ as shown in Fig. \ref{AFBktata}. The behavior of $A_{FB}$ 
is different for $C_7^{eff} > 0$ since it slightly decreases when 
$\bar{\xi}_{N,\tau\tau}^{D}$ increases. In addition to this, $A_{FB}$ is 
sensitive the ratio $\frac{m_{h_0}}{m_{A_0}}$ for $C_7^{eff} > 0$ and 
increases as a magnitude with the decreasing ratio (see Fig. \ref{AFBh0A0}). 
However for $C_7^{eff} < 0$ $A_{FB}$ is not sensitive to the ratio 
$\frac{m_{h_0}}{m_{A_0}}$. Further, it has negative sign for  
both $C_7^{eff} > 0$ and $C_7^{eff} < 0$. 

\begin{figure}[htb]
\vskip -3.0truein
\centering
\epsfxsize=6.8in
\leavevmode\epsffile{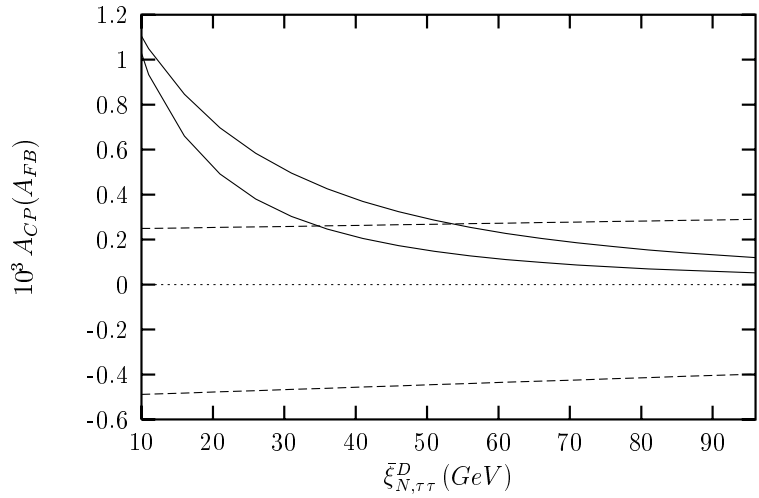}
\vskip -3.0truein
\caption[]{The same as Fig. \ref{Brktata}, but for $A_{CP}(A_{FB})$ as a 
function of $\bar{\xi}_{N,\tau\tau}^{D}$ and $sin\theta=0.5$}
\label{ACPAFBktata}
\end{figure}

\begin{figure}[htb]
\vskip -3.0truein
\centering
\epsfxsize=6.8in
\leavevmode\epsffile{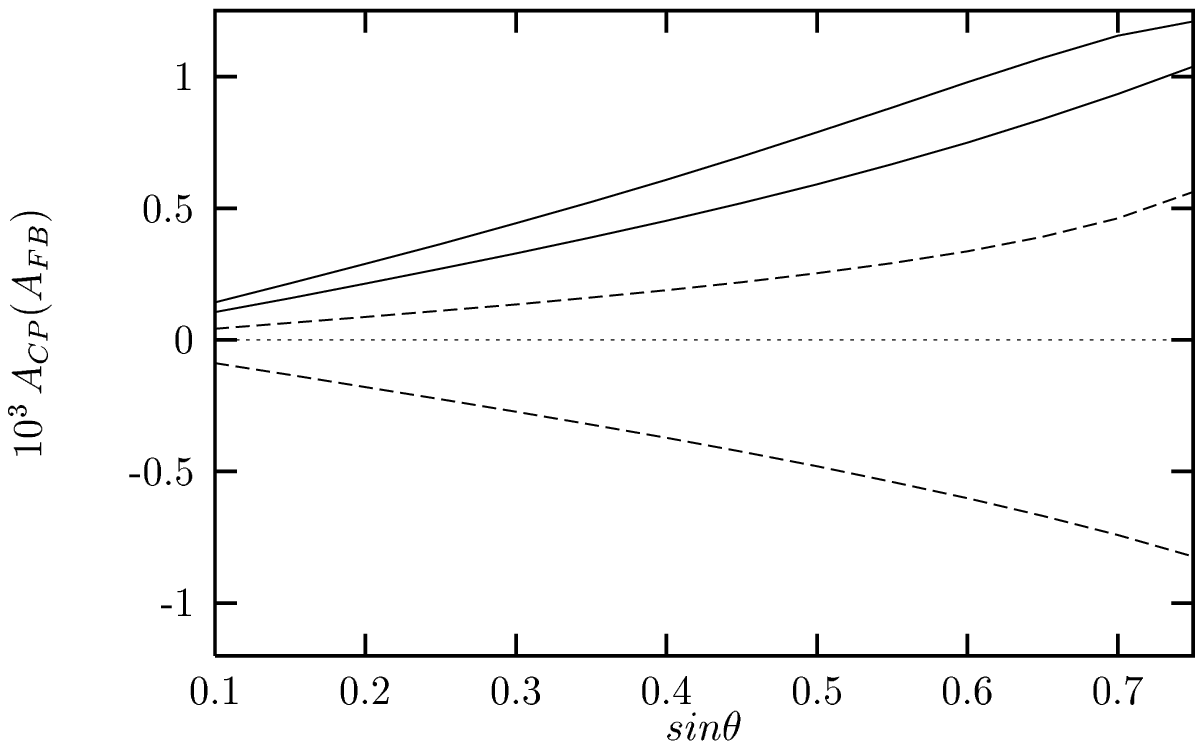}
\vskip -3.0truein
\caption[]{The same as Fig. \ref{ACPsin}, but for $A_{CP}(A_{FB})$ as a 
function of $sin\theta$.}
\label{ACPAFBsin}
\end{figure}

Finally, we present the $CP$ violating asymmetry in $A_{FB}$ 
($A_{CP}(A_{FB})$) in Figs. \ref{ACPAFBktata} and \ref{ACPAFBsin} since 
this parameter exists when NHB effects are non-zero and can play an 
important role in the determination of those effects. 
Fig. \ref{ACPAFBktata} shows $\bar{\xi}_{N,\tau\tau}^{D}$ dependence of 
$A_{CP}(A_{FB})$ for $C_7^{eff} > 0$ (lies between solid lines) and 
$C_7^{eff} < 0$ (lies between dashed lines). This quantity is not so much 
sensitive to $\bar{\xi}_{N,\tau\tau}^{D}$ and can be at the order of 
the magnitude $10^{-4}$  for $C_7^{eff} < 0$. It can have both signs or can 
vanish for this case. For $C_7^{eff} > 0$, the numerical value of 
$A_{CP}(A_{FB})$ can increase up to $10^{-3}$. Here, the SM Higgs $H_0$ 
part of the NHB effects causes to have large values for $A_{CP}(A_{FB})$ 
and the part which contains neutral Higgs bosons beyond enters into 
expression destructively.  Fig. \ref{ACPAFBsin} represents $sin\theta$ 
dependence of $A_{CP}(A_{FB})$. As shown in this figure, the possible 
values of $A_{CP}(A_{FB})$ for $C_7^{eff} > 0$ have the same sign and they
are non-zero for nonzero $sin\theta$, however for $C_7^{eff} < 0$ 
$A_{CP}(A_{FB})$ can vanish or exist with both signs.

Now we would like to present our conclusions.
\begin{itemize}
\item The $Br$ of the exclusive process $B\rightarrow K \tau^+\tau^-$
is at the order of the magnitude $10^{-7}$ for the SM and model III
withouth the NHB effects. However, including the NHB effects and taking large
values of the neutral coupling $\bar{\xi}_{N,\tau\tau}^{D}$, it is possible
to enhance the $Br$ more than one orders of magnitude compared to the one
calculated in the SM. 
\item It would be possible to predict the upper limit of the 
coupling $\bar{\xi}_{N,\tau\tau}^{D}$, $\bar{\xi}_{N,\tau\tau}^{D}< 20\, GeV$, 
respecting the experimental upper limit, $Br ( B\rightarrow K l^+ l^-) < 0.5 
\times 10^{-6}$ , $(l=e,\mu)$ \cite{Babar2}.
\item Using the complex Yukawa coupling $\bar{\xi}_{N,bb}^{D}$ it is
possible to get a $CP$ violating asymmetry $A_{CP}$ at the order of the 
magnitude $10^{-3}$, which is a measurable quantity. With the addition 
of NHB effects the magnitude of $A_{CP}$ decreases.
\item $A_{FB}$ is another physical quantity which exists when the NHB
effects are non-zero. The calculations show that $A_{FB}$ is at the order 
of the magnitude $10^{-2}$ and the experimental measurement of this quantity 
can give strong evidence about the existence of NHB effects and the physics 
beyond the SM.
\item Finally, the $CP$ asymmetry in $A_{FB}$ can appear also with the NHB
effects and it is another physical quantity which can be used for testing 
the existence of the NHB effects. We calculate this quantity at the order 
of the magnitude $10^{-3}$ and its experimental measurement can give 
important clues about physics beyond the SM
\end{itemize}

Therefore, experimental investigations of these physical quantities 
ensure a crucial test for the new physics beyond the SM.
\section{Acknowledgement}
This work was supported by Turkish Academy of Sciences (TUBA/GEBIP).
\newpage
{\bf \LARGE {Appendix}} \\
\begin{appendix}

\section{\bf The model III and the inclusive $b\rightarrow s \tau^+ \tau^-$ 
decay}
In the SM and model I and II 2HDM, the flavour changing neutral current at 
tree level is forbidden. However, such currents are permitted in the general 
2HDM, so called model III and it brings new parameters, i.e. Yukawa 
couplings, into the theory. These couplings are responsible for the 
interaction of quarks and leptons with gauge bosons, namely, the Yukawa 
interaction and in this general case it  reads as 
\begin{eqnarray}
{\cal{L}}_{Y}&=&\eta^{U}_{ij} \bar{Q}_{i L} \tilde{\phi_{1}} U_{j R}+
\eta^{D}_{ij} \bar{Q}_{i L} \phi_{1} D_{j R}+
\xi^{U}_{ij} \bar{Q}_{i L} \tilde{\phi_{2}} U_{j R}+
\xi^{D}_{ij} \bar{Q}_{i L} \phi_{2} D_{j R} \nonumber \\ 
&+& \eta^{D}_{ij} \bar{l}_{i L} \phi_{1} E_{j R}+
\xi^{D}_{ij} \bar{l}_{i L} \phi_{2} E_{j R} +
h.c. \,\,\, ,
\label{lagrangian}
\end{eqnarray}
where $L$ and $R$ denote chiral projections $L(R)=1/2 (1\mp \gamma_5)$,
$\phi_{k}$, for $k=1,2$, are the two scalar doublets, $Q_{i L}$ 
($l_{i L}$) are quark (lepton) doublets, $U_{j R}$ and $D_{j R}$ are 
quark singlets, $E_{j R}$ are lepton singlets, $\eta^{U,D}_{ij}$, 
and $\xi^{U,D}_{ij}$ are the matrices of the Yukawa couplings. 
The Flavor changing (FC) part of the interaction is given by
\begin{eqnarray}
{\cal{L}}_{Y,FC}=
\xi^{U}_{ij} \bar{Q}_{i L} \tilde{\phi_{2}} U_{j R}+
\xi^{D}_{ij} \bar{Q}_{i L} \phi_{2} D_{j R} +
\xi^{D}_{ij} \bar{l}_{i L} \phi_{2} E_{j R} + h.c. \,\, .
\label{lagrangianFC}
\end{eqnarray}
With the choice of $\phi_1$ and $\phi_2$
\begin{eqnarray}
\phi_{1}=\frac{1}{\sqrt{2}}\left[\left(\begin{array}{c c} 
0\\v+H_0\end{array}\right)\; + \left(\begin{array}{c c} 
\sqrt{2} \chi^{+}\\ i \chi^{0}\end{array}\right) \right]\, ; 
\phi_{2}=\frac{1}{\sqrt{2}}\left(\begin{array}{c c} 
\sqrt{2} H^{+}\\ H_1+i H_2 \end{array}\right) \,\, .
\label{choice}
\end{eqnarray}
and the vacuum expectation values,  
\begin{eqnarray}
<\phi_{1}>=\frac{1}{\sqrt{2}}\left(\begin{array}{c c} 
0\\v\end{array}\right) \,  \, ; 
<\phi_{2}>=0 \,\, ,
\label{choice2}
\end{eqnarray}
the SM and beyond can be decoupled.
In eq.(\ref{lagrangianFC}) the couplings  $\xi^{U,D}$ for the FC charged 
interactions are 
\begin{eqnarray}
\xi^{U}_{ch}&=& \xi_{neutral} \,\, V_{CKM} \nonumber \,\, ,\\
\xi^{D}_{ch}&=& V_{CKM} \,\, \xi_{neutral} \,\, ,
\label{ksi1} 
\end{eqnarray}
where  $\xi^{U,D}_{neutral}$ 
\footnote{In all next discussion we denote $\xi^{U,D}_{neutral}$ 
as $\xi^{U,D}_{N}$.} 
is defined by the expression
\begin{eqnarray}
\xi^{U,D}_{N}=(V_L^{U,D})^{-1} \xi^{U,D} V_R^{U,D}\,\, .
\label{ksineut}
\end{eqnarray}
Here the charged couplings appear as linear combinations of neutral 
couplings multiplied by $V_{CKM}$ matrix elements (see \cite{atwood} for
details). 

Now, we would like to present the procedure to calculate the matrix element 
for the inclusive $b\rightarrow s \tau^+ \tau^-$  decay briefly:
\begin{itemize}
\item  Integrating out the heavy degrees of freedom, namely $t$ quark, 
$W^{\pm}$, charged Higgs boson $H^{\pm}$, and neutral Higgs bosons $H_0,\,
H_{1},\,H_{2}$ bosons in the present case and obtaining the 
effective theory. Note that $H_{1}$ and $H_{2}$ are the same as 
the mass eigenstates $h_0$ and $A_0$ in the model III respectively, due 
to the choice given by eq. (\ref{choice}). 
\item Taking into account the QCD corrections through matching the full 
theory with the effective low energy one at the high scale $\mu=m_{W}$ and 
evaluating the Wilson coefficients from $m_{W}$ down to the lower scale 
$\mu\sim O(m_{b})$. 
\end{itemize}
In the 2HDM,  neutral Higgs  particles bring new contributions to the matrix 
element of the process $b\rightarrow s \tau^+ \tau^-$ (see \cite{gurer}) 
since they enter in the expressions with the mass of $\tau$ lepton or related 
Yukawa coupling $\bar{\xi}^{D}_{N,\tau\tau}$. Besides, there exist additional 
operators which are the flipped chirality partners of the former ones 
in the model III. However, the effects of the latter are negligible since 
the corresponding Wilson coefficients are small (see Discussion part). 
Therefore, the effective Hamiltonian relevant for the process 
$b\rightarrow s \tau^+\tau^-$ is
\begin{eqnarray}
{\cal{H}}_{eff}=-4 \frac{G_{F}}{\sqrt{2}} V_{tb} V^{*}_{ts}\aga 
\sum_{i}C_{i}(\mu) O_{i}(\mu)+\sum_{i}C_{Q_i}(\mu) Q_{i}(\mu)\adr 
\, \, ,
\label{hamilton}
\end{eqnarray}
where $O_{i}$ are current-current ($i=1,2$), penguin ($i=3,...,6$),
magnetic penguin ($i=7,8$) and semileptonic ($i=9,10$) operators. Here, 
$C_{i}(\mu)$ are Wilson coefficients normalized at the scale $\mu$. The 
additional operators $Q_{i} (i=1,..,10)$ are due to the NHB exchange 
diagrams and $C_{Q_i}(\mu)$ are their Wilson coefficients (see 
\cite{gurer} for the existing operators and the corresponding Wilson 
coefficients). 
\section{The form factors and the functions appearing in the expressions}
We parametrize the fuctions $A$, $F_1$ and  $F_2$ as 
\begin{eqnarray}
A&=&A_1 + i\, A_2 + \bar{\xi}^D_{N,bb}\, A_3 
\nonumber \,\, \\
F_1&=&F_1^{(1)} + \bar{\xi}^D_{N,bb}\, F_1^{(2)} 
\nonumber \,\, \\
F_2&=&F_2^{(1)} + \bar{\xi}^D_{N,bb}\, F_2^{(2)} 
\label{AF1F2a}
\end{eqnarray}
with 
\begin{eqnarray}
A_1&=&2\, Re(C_9^{eff}) f^+ - \frac{4 m_b f_T}{m_B+m_K} 
C_7^{eff}|_{\bar{\xi}^D_{N,bb}\rightarrow 0} \nonumber \,\, , \\
A_2&=&2\, Im(C_9^{eff}) f^+ \nonumber \,\, , \\
A_3&=& - \frac{4 m_b f_T}{m_B+m_K} \frac{1}{m_b\,m_t} \bar{\xi}^U_{N,tt}
(\eta^{\frac{16}{23}} K_2 (y_t)+\frac{8}{3}
(\eta^{\frac{14}{23}}-\eta^{\frac{16}{23}}) G_2 (y_t))
\nonumber \,\, , \\
F_1^{(1)}&=&\eta^{-12/23}\,\frac{(m_B^2-m_K^2) f^+ + m_B^2 s f^-}{m_b} 
\int^{1}_{0}dx \int^{1-x}_{0} dy \,
(C^{H_0}_{Q_{1}}((\bar{\xi}^{U}_{N,tt})^{2})+
C^{H_0}_{Q_{1}}(\bar{\xi}^{U}_{N,tt})
\nonumber \\ &+&
C^{H_0}_{Q_{1}}(g^{4})+C^{h_0}_{Q_{1}}((\bar{\xi}^{U}_{N,tt})^{3}) 
+ C^{h_0}_{Q_{1}}((\bar{\xi}^{U}_{N,tt})^{2})+
C^{h_0}_{Q_{1}}(\bar{\xi}^{U}_{N,tt}) \nonumber \,\, , \\
F_1^{(2)}&=& \eta^{-12/23}\, \frac{(m_B^2-m_K^2) f^+ + m_B^2 s f^-}
{m_b \bar{\xi}^{D}_{N,bb}} \int^{1}_{0}dx \int^{1-x}_{0} dy 
\,C^{h_0}_{Q_{1}}(\bar{\xi}^{D}_{N,bb}))
\nonumber \,\, , \\
F_2^{(1)}&=&\eta^{-12/23}\,\frac{(m_B^2-m_K^2) f^+ + m_B^2 s f^-}{m_b}
\int^{1}_{0}dx \int^{1-x}_{0} dy\,
(C^{A_0}_{Q_{2}}((\bar{\xi}^{U}_{N,tt})^{3})+
C^{A_0}_{Q_{2}}((\bar{\xi}^{U}_{N,tt})^{2})
\nonumber \\ &+&
C^{A_0}_{Q_{2}}(\bar{\xi}^{U}_{N,tt}) )
\nonumber \,\, , \\
F_2^{(2)}&=&\eta^{-12/23}\,\frac{(m_B^2-m_K^2) f^+ + m_B^2 s f^-}
{m_b \bar{\xi}^{D}_{N,bb}} \int^{1}_{0}dx \int^{1-x}_{0} dy\,
C^{A_0}_{Q_{2}}(\bar{\xi}^{D}_{N,bb})) \,\, ,
\label{AF1F2b}
\end{eqnarray}
where the formfactors $f^+$, $f^-$ and $f_T$ are calculated in the framework 
of the light cone QCD sum rules and represented in the pole forms as
\cite{AOS}
\begin{eqnarray}
f^+&=&\frac{0.29}{1-\frac{m_B^2 s}{23.7}} \,\, , \nonumber \\
f^-&=&-\frac{0.21}{1-\frac{m_B^2 s}{24.3}} \,\, , \nonumber \\
f^T&=&-\frac{0.31}{1-\frac{m_B^2 s}{23}} \,\, .
\label{formfac}k
\end{eqnarray}

The other functions $B$, $C$ and $D$ appearing in eq. (\ref{matr1}) are  
\begin{eqnarray}
B&=&C_9^{eff} (f^- + f^+) + C_7^{eff}\, \frac{2 m_b f_T}{s} 
\frac{(1-r-s )}{m_B+m_K} \,\, , \nonumber \\
C&=& 2 C_{10} f^+ \,\, , \nonumber \\
D&=& C_{10} (f^- + f^+)  \,\, .
\label{BCD}
\end{eqnarray}
\end{appendix}

\newpage

\begin{thebibliography}{1}
\bibitem{Hewett} J. L. Hewett, in Proc. of the $21^{st}$ Annual SLAC Summer 
Institute, ed. L. De Porcel and C. Dunwoode, SLAC-PUB-6521 (1994)
%
\bibitem{Babar1} The Babar Collaboration,  {\it hep-ex}/0107026. 
%
\bibitem{CLEO2} CLEO Collaboration,
{\it Phys. Rev. Lett.} {\bf 87} (2001) 181803.
%
\bibitem{Belle} The Belle Collaboration,  
{\it Phys. Rev. Lett.} {\bf 88} (2002) 029801.
%
\bibitem{Babar2} The Babar Collaboration, {\it hep-ex}/0201008.
%
\bibitem{Hou} W. -S. Hou, R. S. Willey and A. Soni,
{\it Phys. Rev. Lett.} {\bf 58} (1987) 1608.
%
\bibitem{R5} N. G. Deshpande and J. Trampetic,
{\it Phys. Rev. Lett.} {\bf 60} (1988) 2583.
%
\bibitem{R6} C. S. Lim, T. Morozumi and A. I. Sanda,
{\it Phys. Lett.} {\bf B218} (1989) 343.
%
\bibitem{Grinstein1} B. Grinstein, M. J. Savage and M. B. Wise,
{\it Nucl. Phys.} {\bf B319} (1989) 271.
%
\bibitem{R8} C. Dominguez, N. Paver and Riazuddin, 
{\it Phys. Lett.} {\bf B214} (1988) 459.
%
\bibitem{R9} N. G. Deshpande, J. Trampetic and K. Ponose,
{\it Phys. Rev.} {\bf D39} (1989) 1461.
%
\bibitem{Jaus} W. Jaus and D. Wyler,
{\it Phys. Rev.} {\bf D41} (1990) 3405.
%
\bibitem{donnel} P. J. O'Donnell and H. K. Tung,
{\it Phys. Rev.} {\bf D43} (1991) 2067.
%
\bibitem{R12} N. Paver and Riazuddin,
{\it Phys. Rev.} {\bf D45} (1992) 978.
%
\bibitem{ali} A. Ali, T. Mannel and T. Morozumi,
{\it Phys. Lett.} {\bf B273} (1991) 505.
%
\bibitem{R14} A. Ali, G. F. Giudice and T. Mannel,
{\it Z. Phys.} {\bf C67} (1995) 417.
%
\bibitem{R15} C. Greub, A. Ioannissian and D. Wyler,
{\it Phys. Lett.} {\bf B346} (1995) 145; \\
D. Liu {\it Phys. Lett.} {\bf B346} (1995) 355; \\
G. Burdman, {\it Phys. Rev.} {\bf D52} (1995) 6400: \\
Y. Okada, Y. Shimizu and M. Tanaka {\it Phys. Lett.} {\bf B405} (1997) 297.
%
\bibitem{buras} A. J. Buras and M. M\"{u}nz,
{\it Phys. Rev.} {\bf D52} (1995) 186.
%
\bibitem{Deshpande} N. G. Deshpande, X. -G. He and J. Trampetic,
{\it Phys. Lett.} {\bf B367} (1996) 362.
%
\bibitem{Geng} C. Q. Geng, C. P Kao, 
{\it Phys. Rev.} {\bf D54} (1996) 5636. 
%
\bibitem{ASOK} T. M. Aliev, A. \"{O}zpineci, M.Savc{\i}, and H. Koru
{\it J. Phys. } {\bf G24} (1998) 49 .
[A%
\bibitem{Demir} D. A. Demir, K. A. Olive and M. B. Voloshin,
{\it hep-ph}/0204119
%
\bibitem{Casalbuoni} R. Casalbuoni, A. Deandra, N. Di Bartolemo, 
R. Gatto and G. Nardulli,\\
{\it Phys. Lett.} {\bf B312} (1993) 315.
%
\bibitem{Colangelo} P. Colangelo, F. De Fazio, P. Santorelli and E. 
Scrimieri, {\it Phys. Rev.} {\bf D53} (1996) 3672.
%
\bibitem{Roberts} W. Roberts,
{\it Phys. Rev.} {\bf D54} (1996) 863.
%
\bibitem{AOS} T. M. Aliev, A. \"{O}zpineci and M.Savc{\i},
{\it Phys. Lett.} {\bf B400} (1997) 194 .
%
\bibitem{gurer} E. Iltan and G. Turan,  
{\it Phys. Rev.} {\bf D63} (2001) 115007.
%
\bibitem{cleo2} M. S. Alam, CLEO Collaboration, to appear in ICHEP98 
Conference (1998)
%
\bibitem{alil2} T. M. Aliev, and E. Iltan, 
{\it Phys. Rev.} {\bf D58} (1998) 095014.
%
\bibitem{atwood} D. Atwood, L. Reina and A. Soni, 
{\it Phys. Rev.} {\bf D55} (1997) 3156.
%
\bibitem{david} D. B. Chao, K. Cheung and W. Y. Keung,
{\it Phys. Rev.} {\bf D59} (1999) 115006.
%
\bibitem{alil1} T. M.Aliev, E. Iltan 
{\it J. Phys. } {\bf G25} (1999) 989.
%

\end{thebibliography}
\end{document}